\documentclass[conference]{IEEEtran}
\IEEEoverridecommandlockouts
\usepackage{cite}
\usepackage{amsmath,amssymb,amsfonts}
\usepackage{algorithmic}
\usepackage{multirow}
\usepackage{hhline}
\usepackage{graphicx}
\usepackage{textcomp}
\def\BibTeX{{\rm B\kern-.05em{\sc i\kern-.025em b}\kern-.08em
    T\kern-.1667em\lower.7ex\hbox{E}\kern-.125emX}}
\begin{document}

\title{Data Mining in Scientometrics: usage analysis for academic publications}

\author{\IEEEauthorblockN{Olesya Mryglod}
\IEEEauthorblockA{\emph{Laboratory for Statistical Physics of Complex Systems} \\
\emph{Institute for Condensed Matter Physics, NAS of Ukraine} \\ Lviv, Ukraine\\
\textbf{L}$^4$ Collaboration \& Doctoral College
for the Statistical Physics of Complex Systems,\\
Leipzig-Lorraine-Lviv-Coventry, Europe
             olesya@icmp.lviv.ua}
\and
\IEEEauthorblockN{Yurij Holovatch}
\IEEEauthorblockA{\emph{Laboratory for Statistical Physics of Complex Systems} \\
\emph{Institute for Condensed Matter Physics, NAS of Ukraine} \\ Lviv, Ukraine\\
\textbf{L}$^4$ Collaboration \& Doctoral College
for the Statistical Physics of Complex Systems,\\
Leipzig-Lorraine-Lviv-Coventry, Europe}
\and
\IEEEauthorblockN{Ralph Kenna}
\IEEEauthorblockA{\emph{Applied Mathematics Research Centre, Coventry University}\\
              Coventry, CV1 5FB, England\\
\textbf{L}$^4$ Collaboration \& Doctoral College
for the Statistical Physics of Complex Systems,\\
Leipzig-Lorraine-Lviv-Coventry, Europe}
}
\maketitle

\begin{abstract}
We perform a statistical analysis of scientific-publication data with a goal to provide quantitative analysis of scientific process. Such an investigation belongs to the newly established field
of scientometrics: a branch of the general science of science that covers all
quantitative methods to analyze science and research process. As a case study
we consider download and citation statistics of the journal ``Europhysics Letters'' (EPL),
as Europe's flagship letters journal of broad interest to the physics community.
While citations are usually considered as an indicator of academic impact,
downloads reflect rather the level of attractiveness or popularity of a publication.
We discuss peculiarities of both processes and correlations between them.
\end{abstract}

\begin{IEEEkeywords}
scientometrics, usage metrics, citation analysis, data analysis
\end{IEEEkeywords}

\section{Introduction}
Research, as any other kind of human activity, can be considered as a source for large sets or streams of multidimensional data, starting from trivial
statistics of the number of researchers or the number of papers they produce \cite{Nalimov71,Price63} and ending with records of communication through social networks on special
topics in real time \cite{Priem2014}. The opportunity to record, to store, and to analyze scientific-related data itself has become a trigger for the extremely fast development
of scientometrics \cite{Scientometrics_review2015} --- a branch of the general science of science that covers all quantitative methods to analyze science and research processes.
The majority of scientometric tasks are highly applicable in practice. This is especially true today, when the proportion of researchers has become large enough to create a competitive environment and to become an object of careful review by governments
\cite{Price63}. It is always good to present some new knowledge about the universe or some phenomenological findings, but nowadays the researcher is facing more mercantile questions such as: what is the practical impact, how the new results can be implemented into the technology, etc. Therefore, numerous quantitative indicators and metrics related to science have become candidates for immediate implementation in assessment procedures on different scales. This is not bad thing in general, but the important caveats should be kept in mind: (i) it is extremely hard to formalize any human activity in general, (ii) the work of scientist is not mechanical, but to large extent it is a creative process and (iii) science is far from being homogeneous --- different disciplines should be evaluated in completely different ways. While the controlling and assessment of research activity seems to be the requirement of time, it should be performed in a non-trivial way and taking into account the multidimensionality of data used.

While journal publication remains the most common way to present the results of research, starting from the middle of past century citations became a
sort of `currency' in science \cite{Garfield79}. Citation data are currently used for calculation of different indicators of research efficiency despite numerous drawbacks and caveats. Even not considering the peculiarities of citation counting, the meaning of a single citation is ambiguous. The motivation to cite a particular source can be various \cite{Tahamtan18}. Therefore, recently it is usually considered as the unit of \emph{impact}. The impact can be positive or negative; it can represent the usefulness of a methodology, the novelty of subject or the informativeness of a review presented in a publication; it can be a form of acknowledgement; and so on.

The increasing popularity of different virtual social media, blogs and services such as managers of references, bookmarking, etc. caused the appearance of new source for scientometric data. New categories of metrics, called altmetrics or informetrics, were presented some time ago \cite{Priem2014}. The latter obviously contain completely different information about publications.
At this point one can think about such special characteristics as popularity. This can be already dependent not solely on scientific value of results, but also on the way how they are presented and promoted.
One of the motivations behind is that even a good result often should be well positioned in such competitive environments. Even a very special `drop in an ocean' can remain unseen, not to make an impact, to stay uncited and, therefore, to be underestimated. There are the kind of data which can say something about the attractiveness of the paper even before it is read. The statistics of downloads are referred to as
one of the so-called usage metrics. Even if downloading never guarantees that the paper will be fully or even partially read, it can be considered as the unit of reader's interest.

Considering downloads and citations --- two dimensions of data characterizing research papers --- one would initially speculate about the correlation between them. On the one hand, it is intuitively acceptable
that, in the main, the wider the readership (downloads), the greater the probability that a document will be cited (citations). On the other hand, it has already been shown that downloads may influence citations and vise versa \cite{Moed2005,Moed2016}. The correlation coefficients between downloads and citations reported in the literature seem to be different not only for different disciplines, but for different journals and even for different types of publications \cite{Moed2016}.
This brings us to the clustering problem - the papers (journals?) characterized by similar patterns in downloading and citing should be grouped and analyzed separately in order to understand the nature of both processes and the relationship between them. In this paper we discuss a case study containing the results of the analysis of statistics of full-text download and corresponding citations of
publications in \emph{EPL} (formerly known as ``Europhysics Letters'') ---
Europe's flagship peer-reviewed letters journal of broad interest to the physics community. It is  published by EDP Sciences,
IOP Publishing and the Italian Physical Society on behalf of the European Physical Society and 17 other European physical societies \cite{EPLweb,WikiEPL}.
The rest of the paper is organized as follows:  \emph{EPL}  download statistics is analysed in the next Section \ref{II} and is compared
with the citation statistics in Section \ref{III}. We end by conclusions and outlook in Section \ref{IV}.

\section{Evolution of download process} \label{II}

Scientific citations have been an object of careful consideration starting from the middle of past century.
Even if collecting the citations is a highly non-trivial process, there are few comparatively reliable sources of the corresponding data (Web of Science \cite{WoS}, Scopus \cite{Scopus}). The downloads, on the other hand, have become a topic of interest later \cite{Kurtz2005}. Download statistics are not publicly available --- these have to be provided by publishers or, say, by owners of repositories. Therefore, the analysis of downloading process is important itself as well.

What is the downloading `profile' of any scientific periodical? Does some typical downloading pattern for individual paper exist? Is it possible to perform clustering of publications according to their attractiveness for readers? Some clues to answer these questions can be found in this section containing the results of analysis of downloads for \emph{EPL} \cite{EPL}. The data on the full-text downloads for papers, published  in {\emph{EPL}} between January 2007  to June 2013 with one month resolution provided by \emph{IOPscience}\footnote{\emph{IOPscience} is the online service for the journals of the \emph{Institute of Physics (IoP)}} were used for this purpose. The downloads are counted on an IP-address basis with multiple requests made from the same address considered as separate downloads. Only full-text downloads from the {\emph{IOPscience}} web-pages are counted. The data are automatically cleansed of suspicious and robot activity, and are  COUNTER compliant, see \cite{COUNTER}. The data set comprises 377 open access (OA) papers (freely available via web-page for unregistered users) and 4\,986 non-OA papers (require a payment).

\begin{figure}[htbp]
\centerline{\includegraphics[width=0.5\textwidth]{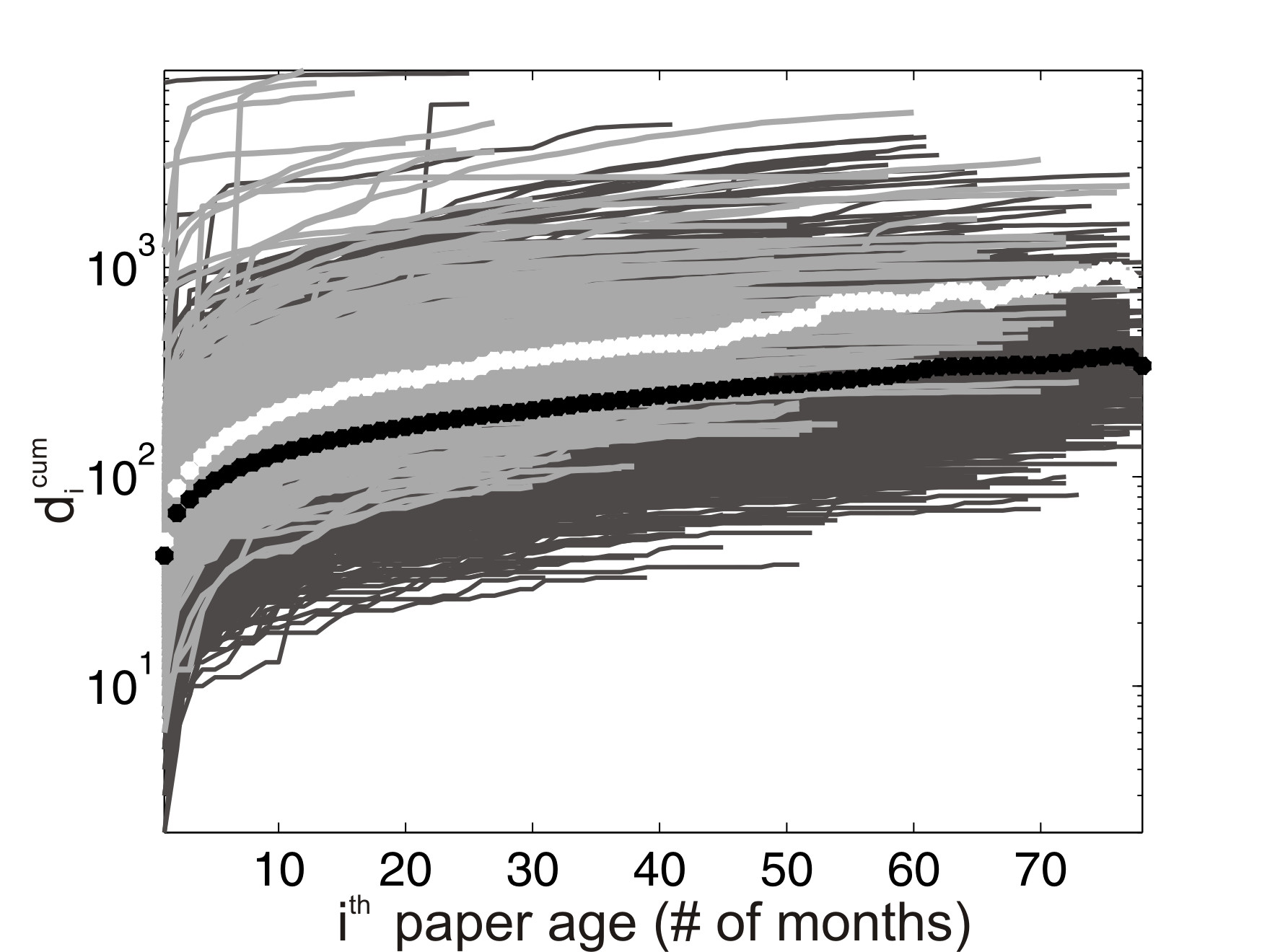}}
\caption{Cumulative number of downloads for each paper individually vs. its age online: dark lines represent non-OA papers while light lines represent OA papers. The corresponding median
values for papers of the same age are indicated by the circles (black for non-OA papers and white for OA papers). }
\label{fig1_diach_all}
\end{figure}

We used two approaches to analyse the time-involving sequences of data: (i) to fix the period of real time (say, a specific month, year, etc.) and to analyse the corresponding downloading statistics for papers of different ages\footnote{Here the paper's age is calculated as the number of months passed since it has been published on the web-page.} --- the so-called \emph{synchronous} method; or (ii) to consider the statistics of downloads of individual paper in its personal timescale, which starts at paper's publication online  --- the \emph{diachronous} approach. The latter allows one to compare the individual processes for separate publications in order to find patterns. Cumulative numbers of downloads $d_i^{\mathrm{cum}}$ vs. corresponding paper's ages are plotted in Fig.~\ref{fig1_diach_all}. Each curve here represent a given paper and the value $(x,y)$ means that this paper has got $y$ downloads during $x$ months after its publication.

Having the family of curves, there is a temptation to get some typical curve by averaging all values of $d^{\mathrm{sum}}_i$ for each age. Since the distribution of these values is skewed, median instead of averages should rather be used, see Fig.~\ref{fig1_diach_all}. The root-mean-square deviation (RMSD) of $d^{cum}_i$  from the median values can be used to gauge the extent to which the latter can be considered as typical ones. The value $\mathrm{RMSD_c}=66$ is empirically determined as critical for non-OA papers ($\mathrm{RMSD_c}=105$ for OA papers) to detect the individual curves most close to median values. This allows us to determine the core of typical papers: about 60\% of non-OA and 50\% of OA papers. The fast accumulation of downloads seems to be typical during the first couple of months after publication, while the process is getting slower afterwards. The rest of the papers --- let us call them atypical to distinguish from the first group --- are basically characterized by the curves of a similar smooth shape, but different  rates of downloads accumulation. However, some exceptions can be visually seen in Fig.~\ref{fig1_diach_all}: unusual bursts can be noticed for particular papers. In order to capture such bursty behaviour the standard deviations can be used. Let $\sigma_i (T)$ be the standard deviation of $i$th paper over its entire history to month $T$. The median values are considered as benchmark to be compared to. $\sigma(T)$ is the standard deviation calculated for median values but only over the first $T$ months. Each paper is then characterized by a value of $\Delta_i(T)$:
\[
\Delta_i(T)=|\sigma_i(T)-\sigma(T)|.
\]
The noticeable bursts can be flagged by large values of $\Delta_i(T)$ comparing to the average $\langle \Delta_i(T)\rangle$.  The investigation of ratio $\Delta_i(T)/\langle \Delta_i(T)\rangle$ lead us to its critical value equals to 5. In this way, 96 non-OA papers (2\% of all) and 10 OA papers (2.7\%) demonstrating noticeable bursts were detected. The position and the nature of bursts can the issue of separate analysis. E.g., some of bursty papers can be labelled as `sleeping beauties' due to its delayed recognition (the downloading bursts occurred later than 6 months after publication for 35 non-OA and for 2 OA papers), see Table~\ref{tab1}.

Continuing with atypical non-bursty papers, we can also analyse their overall attractiveness. The cumulative curves of the more than 1/3 of these papers lie entirely over the median values. Another 1/3 of them are of persistently lower attractiveness comparing to medians. The rest are characterized by cumulative curves which cross medians curve, see Table~\ref{tab1}.

Another way to investigate the downloading rates for non-bursty papers is to use the notion of half-life $M^{50}_i$ which is the number of months by which a paper achieve 50\% of its current downloads. If this value falls between the percentiles P25 and P50, the corresponding paper is considered as usual, i.e. demonstrating the normal typical rate of downloads accumulation. The $M^{50}_i$ value smaller than P25 means that 50\% of downloads were acquired by $i$th paper faster than usual --- we called such papers `flashes in the pans'. In contrast, the `delayed' papers are characterized by $M^{50}_i$ values greater than P50. The exact numbers of \emph{EPL} publications within these categories are shown in Table~\ref{tab1}.

\begin{table}[!h]
\vspace{-5mm}
\label{tab1}
\caption{Categorisation of {\em{EPL}}  papers according to downloads (the data for OA papers are in brackets)  \cite{EPL}.
}
\vspace{2ex}
\footnotesize{
\begin{tabular}{|l|l|l|}
\hline
\multicolumn{3}{|c|}{\parbox[t]{8cm}{\textbf{Categorisation by burstiness}}}\\
\hline
\multirow{3}{*}{\parbox[t]{1.5cm}{4\,986 (377) papers }} &\multicolumn{2}{|l|}{ 98\% are ``non-bursty'' papers }\\
 \hhline{~--}
&\multirow{2}{*}{\parbox[t]{1.8cm}{2\% (3\%) are ``bursty'' papers}}&\parbox[t]{4.25cm}{1\% (1\%) are ``sleeping beauties''}\\
\cline{3-3}
& & 1\% (2\%) burst early  \\
\hline\hline
\multicolumn{3}{|c|}{\parbox[t]{8cm}{\textbf{Categorisation by overall attractiveness}}}\\
\hline
\multirow{3}{*}{\parbox[t]{1.5cm}{4\,890 {(367)} non-bursty papers}} &\multicolumn{2}{|l|}{{60\%} {(50\%)} {have typical overall attractiveness}}\\
\hhline{~--}
&\multirow{3}{*}{\parbox[t]{1.6cm}{40\%{(50\%)} are atypical}}%
&\parbox[t]{4.45cm}{{18\%} {(22\%)} are more attractive \vspace{0.8ex}}\\
\cline{3-3}
&&\parbox[t]{4.45cm}{12\% {(15\%)} are less attractive \vspace{0.8ex}}\\
\cline{3-3}
&&\parbox[t]{4.45cm}{10\% {(13\%)} the rest \vspace{0.8ex}}\\
\hline\hline
\multicolumn{3}{|c|}{\parbox[t]{8cm}{\textbf{Categorisation by half-lives (ageing of attractiveness)}}}\\
\hline
\multirow{3}{*}{\parbox[t]{1.5cm}{4\,890 {(367)} non-bursty papers}} &\multicolumn{2}{|l|}{ 62\% {(65\%)}  {exhibit usual ageing behaviour}}\\
 \hhline{~--}
&\multicolumn{2}{|l|}{ {18\%  (17.5\%)} are flashes-in-the-pan}\\
\hhline{~--}
 &\multicolumn{2}{|l|}{ {20\% (17.5\%)} exhibit delayed activity}\\
\hline
\end{tabular}}
\end{table}

\begin{figure}[htbp]
\centerline{\includegraphics[width=0.5\textwidth]{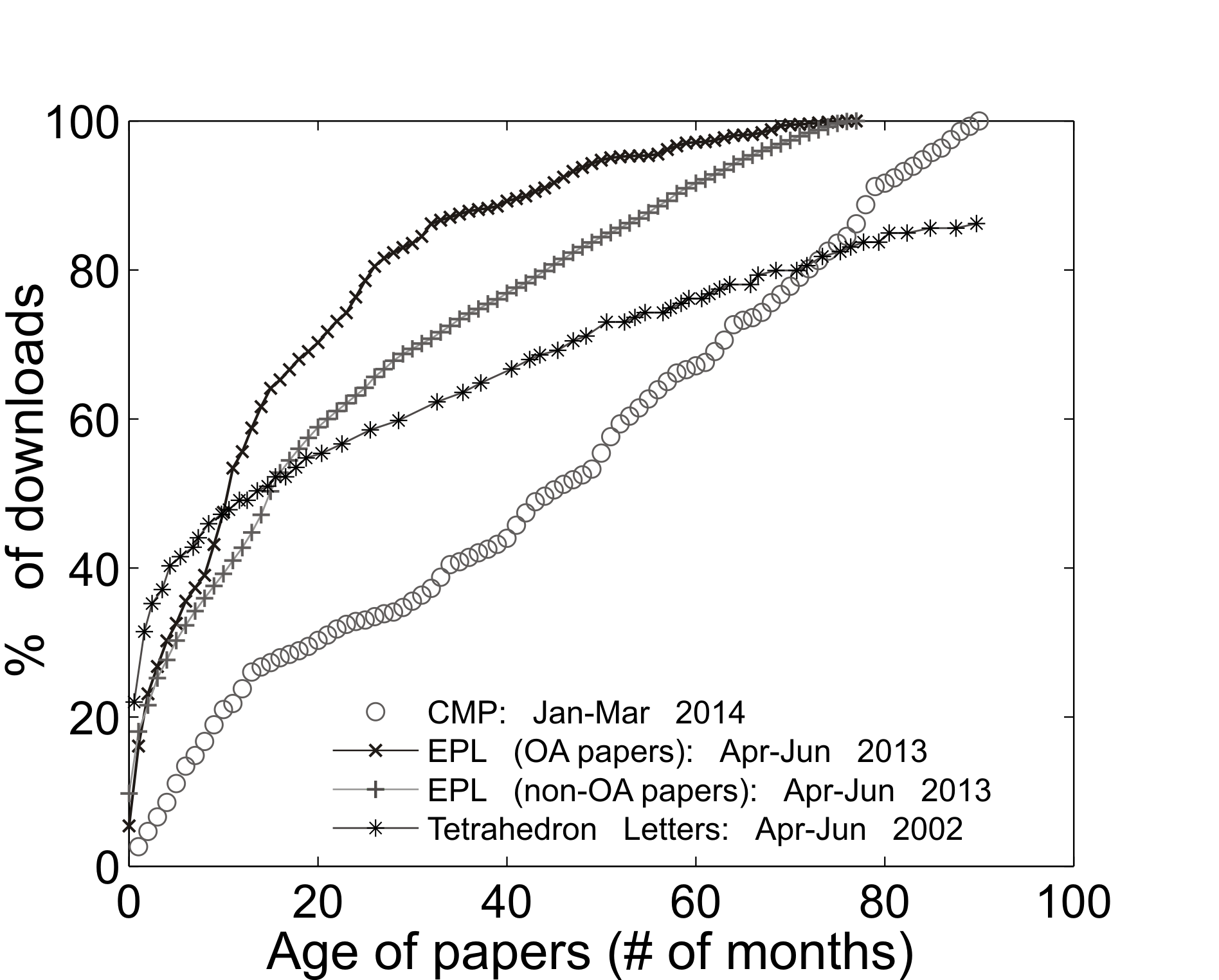}}
\caption{Relative cumulative distributions of ages of downloaded papers. The ``+'' symbols represent
non-OA \emph{EPL} papers, ``x'' represent OA \emph{EPL} papers \cite{EPL}, circles correspond to CMP data, and ``$\ast$'' represent data for Tetrahedron Letters~\cite{Moed2005}.}
\label{fig2_sync}
\end{figure}

Coming back to \emph{synchronous} approach, we can get the downloading patterns of the entire journal on a calendar basis --- i.e., for different time periods. In this way 3-months observation periods were chosen to compare the statistics of downloads versus paper ages for three journals: \emph{EPL} (OA and non-OA papers), \emph{Tetrahedron Letters} and \emph{Condensed Matter Physics}. The second curve is the digitized version of Fig.~1 from ref.~\cite{Moed2005} which corresponds to download data for the journal \emph{Tetrahedron Letters}. The third one is based on the publisher's downloading data containing 1201 records with one month resolution. Each symbol $(x,y)$ in Fig.~\ref{fig2_sync} means that $y\%$ of all downloads accumulated within given time period were to papers not older than $x$ months.

\begin{figure}[htbp]
\centerline{\includegraphics[width=0.5\textwidth]{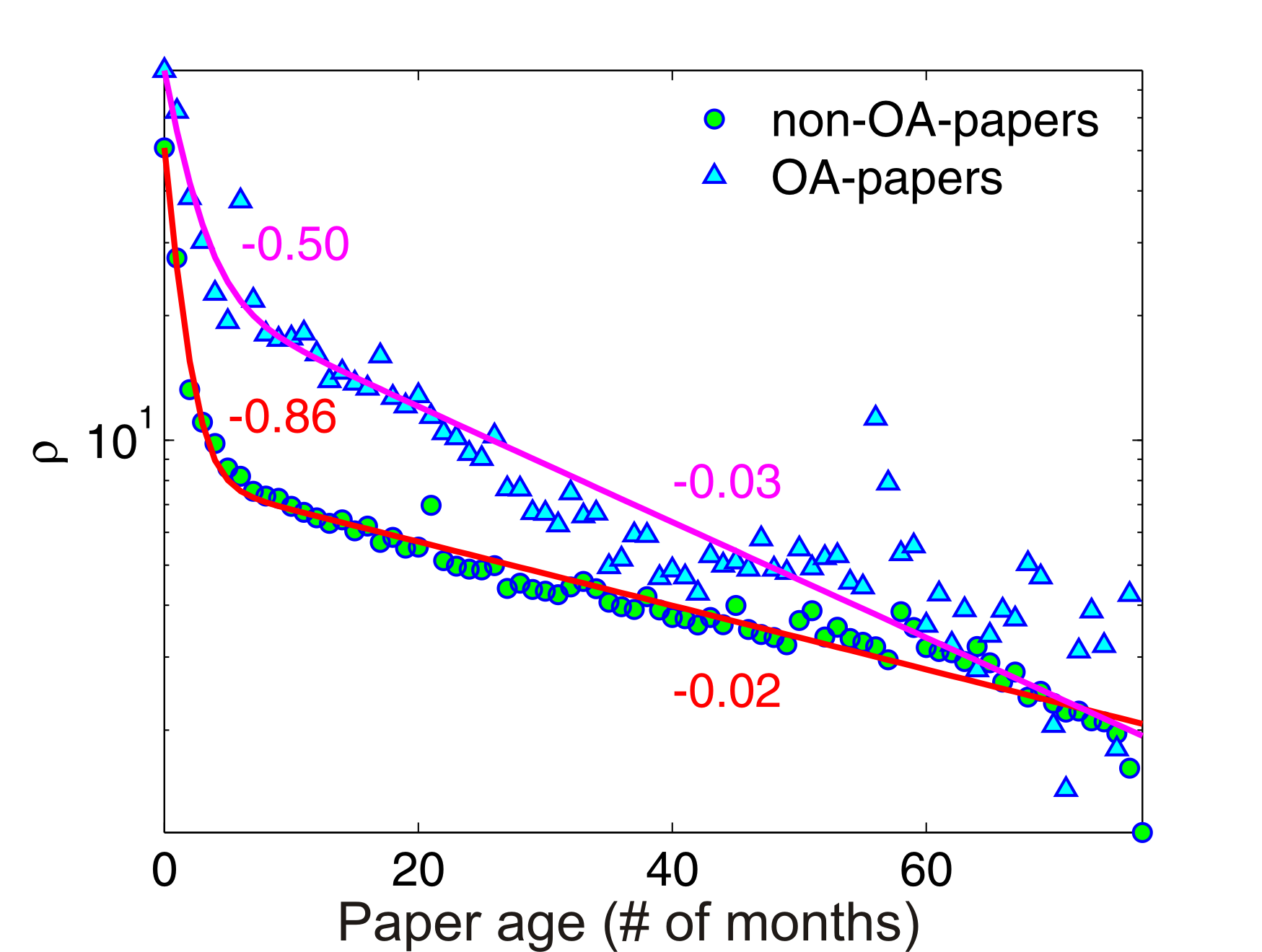}}
\caption{Density of downloads per paper ($\rho$) versus papers' ages.
The solid curves show the model (\ref{model}) predictions and the corresponding exponential decay constants are indicated.}
\label{fig_model}
\end{figure}

As one can see in Fig.~\ref{fig2_sync}, the downloading process looks different for all three journals. This is rather an expected finding since they differ by discipline (Physics for \emph{EPL} and \emph{CMP}, Organic Chemistry for \emph{Tetrahedron Letters}), dominant kind of publications (rapid short publications in \emph{EPL} and \emph{Tetrahedron Letters}, and original regular papers in \emph{CMP}), frequency (\emph{Tetrahedron Letters} publishes one issue per week; \emph{EPL} --- per two weeks; \emph{CMP} --- per three months) and the rules for online access (full open access in \emph{CMP} and hybrid scheme in the rest two). However, careful consideration reveals similar features: a rather faster accumulation of downloads during couple of months immediately after publication online and further slowering the process afterwards. In ref.~\cite{Moed2005} a critical point of three months was discussed, as a demarcation between two different regimes characterised by different slopes (see also \cite{Watson2009}). A two-factor model was proposed in \cite{Moed2005} to describe the corresponding two regions which supposedly can be explained by two initial motivations of users: to download more recent papers because of their novelty and older ones for archiving, background reading or similar. According to this model, two exponents can be used in order to fit non-cumulative data (we plot the density of downloads $\rho$ -- defined as  mean numbers of downloads per paper -- against their age see Fig.~\ref{fig_model}):
\begin{eqnarray}
\label{model}
\rho(t)=\rho_0\left[ A\exp(-b_1 t)+ (1-A)\exp(-b_2 t)\right], \\
\nonumber 0\leq A\leq1, \qquad b_1>0, \qquad b_2>0\,,
\end{eqnarray}
where $A$ and $(1-A)$ are relative weights of the two factors (two different motives for downloads) and $\rho_0$ is the density of downloads which corresponds to the newest papers (published in the month of downloading). The parameters $b_1$ and $b_2$ are  exponential decay constants corresponding to early and later download patterns.

Using  nonlinear-curve least-squares fitting, we obtain the estimates for non-OA papers $A \approx 0.84$, $b_1 \approx 0.86$, $b_2 \approx 0.02$ and for OA papers $A \approx  0.71$, $b_1 \approx 0.50$, $b_2 \approx 0.03$ (parameters close to the estimates of \cite{Moed2005} for {\emph{Tetrahedron Letters}} which are: $A \approx 0.92$, $b_1 \approx 0.50$, $b_2 \approx 0.014$). The OA downloads are more concentrated on the first months after publication online. The model allows one to make predictions for long-term behaviour: e.g., typically $50\%$ of non-OA downloads collected during by one month are to papers 25 months old or less, see \cite{EPL}.

\section{Downloads vs. citations} \label{III}
Probably the hottest topic of interest is connected with analysis of correlation between downloads and citations. Due to the current applicability of citations, the downloads are usually considered as a potential proxy for future citations. A number of studies were already done in this direction. It seems that such desirable correlations are in fact dependent on different factors such as discipline, type of publications or the peculiarities of a specific journal \cite{Moed2016,Gorraiz2014}. Still, there is a lack of clear understanding of relation between the two processes. While for a sake of statistical reliability aggregated data for several journals considered as a sort of super-journal are
sometimes analysed, some specific features of separate journals can be missed. Therefore, any useful idea towards revealing clusters of journals (papers) with similar downloads-citations patterns is required. This case study is another contribution in this direction.
\begin{figure}[!ht]
\centerline{\includegraphics[width=0.5\textwidth]{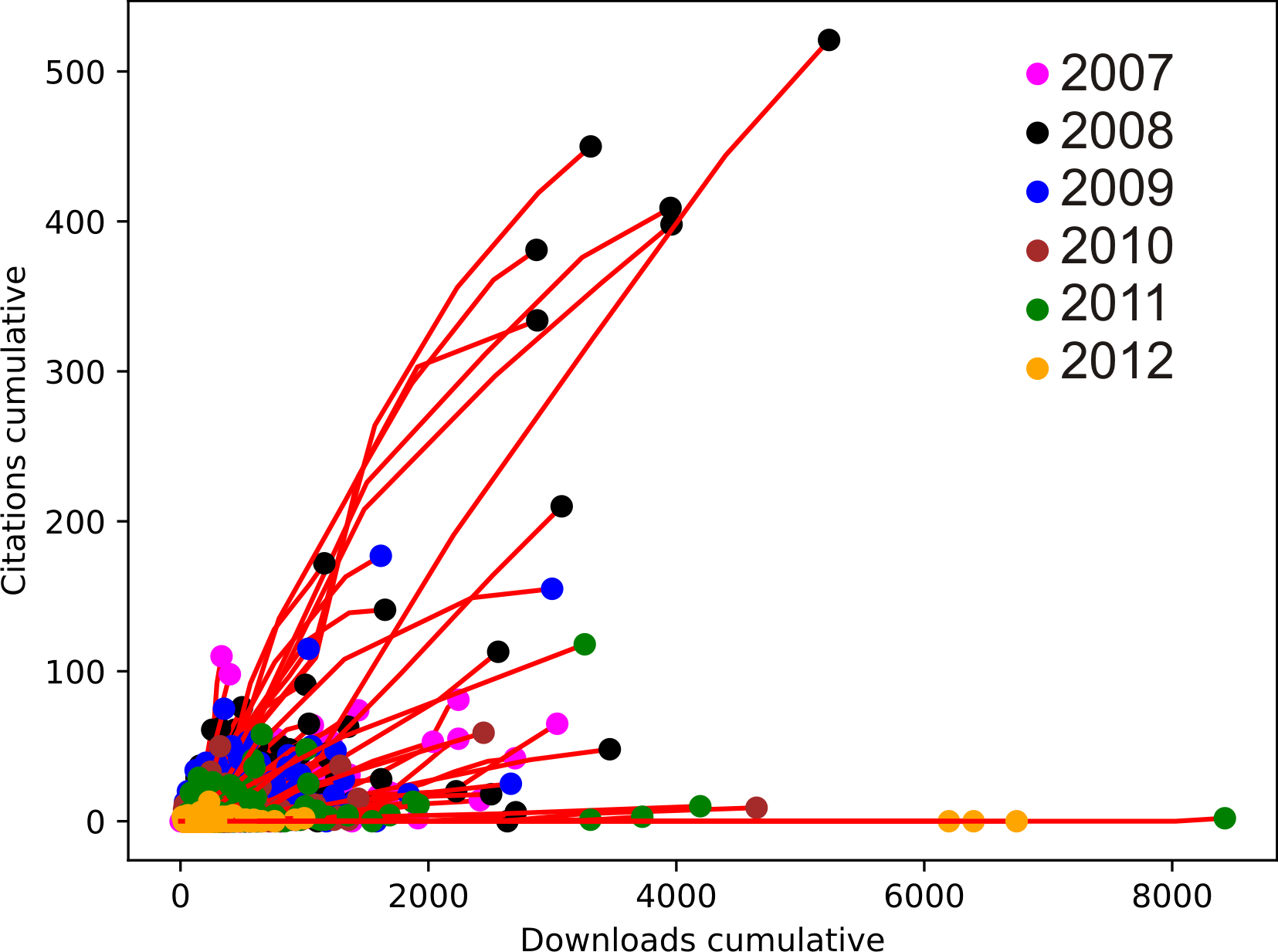}}
\caption{Cumulative downloads vs. cumulative citations. The growing of annual values is shown by line while the final cumulative values are marked by circles. The colors of circles correspond to different publication years.}
\label{fig3_d_vs_c_cum}
\end{figure}

The citation data for 4\,894 \emph{EPL} papers were acquired from the Scopus database. In order to compare it with the existing data, the same publication and citation windows were chosen: [2007--2012]. Therefore, two sequences of values were assigned to the majority\footnote{Some papers cannot be found in Scopus probably for technical reasons.} of \emph{EPL} papers published within 2007 and 2012 years. Since only annual citation data are available, the corresponding downloads were also grouped for years. The annual pairs of cumulative values (downloads vs. citations) per each paper are plotted in Fig.~\ref{fig3_d_vs_c_cum}.

\begin{figure}[!ht]
\centerline{\includegraphics[width=0.5\textwidth]{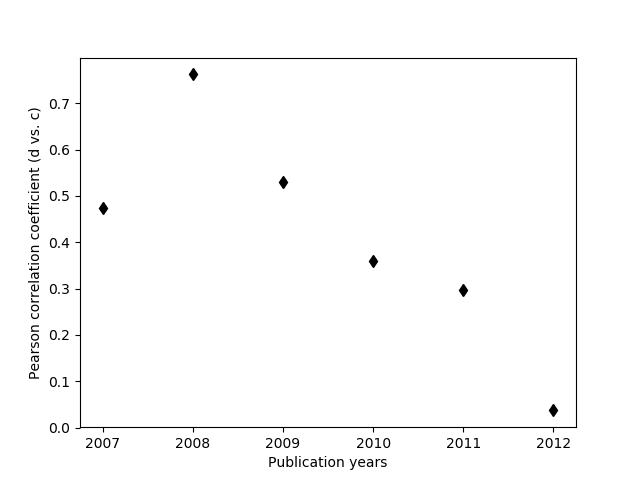}}
\caption{Pearson correlation coefficients (PCC) characterizing the correlation between the cumulative numbers of downloads $d$ and citations $c$ vs. publication year of papers. }
\label{fig_pcc}
\end{figure}

The very natural first step is to find the correlation between the final (cumulative) values of downloads $d$ and citations $c$ shown by circles in Fig.~\ref{fig3_d_vs_c_cum}. Rather low value of Pearson correlation coefficient  $\mathrm{PCC}\approx 0.49$ shows the absence of evident correlation between the for corresponding values for all papers. The correlation coefficient values for papers published in different years are shown in Fig.~\ref{fig_pcc}: higher for papers with longer history and lower for more recent publications. The results are fairly similar for cumulative citations counted in further two years: 2013 and 2014. In spite of the expectations to see better correlation after taking into account 2-years time delay for citations \cite{Wan2010}, the cumulative downloads $d_{2012}$ are not better correlated with $c_{2013}$ and $c_{2014}$.

Further speculations are relevant to the shape of $d$ vs. $c$ cumulative curves (lines in Fig.~\ref{fig3_d_vs_c_cum}). The straight lines are supposed to represent the downloading and citing processes governed by constant motives: the annual increment of ratio $c/d$ remains the same to some extent. The zig-zag-like shape, in the opposite case, demonstrates the change of initial motives providing more attention by readers (faster accumulation of downloads) or more acknowledgement by citing authors (more intensive citing process).

The sequence of value of angles $\alpha$ between the curve and the horizontal axis characterizes the zig-zag-ness of each long enough cumulative curve $d$ vs. $c$, see an example in Fig.~\ref{fig_example_zigzag}.
\begin{figure}[!ht]
\centerline{\includegraphics[width=0.5\textwidth]{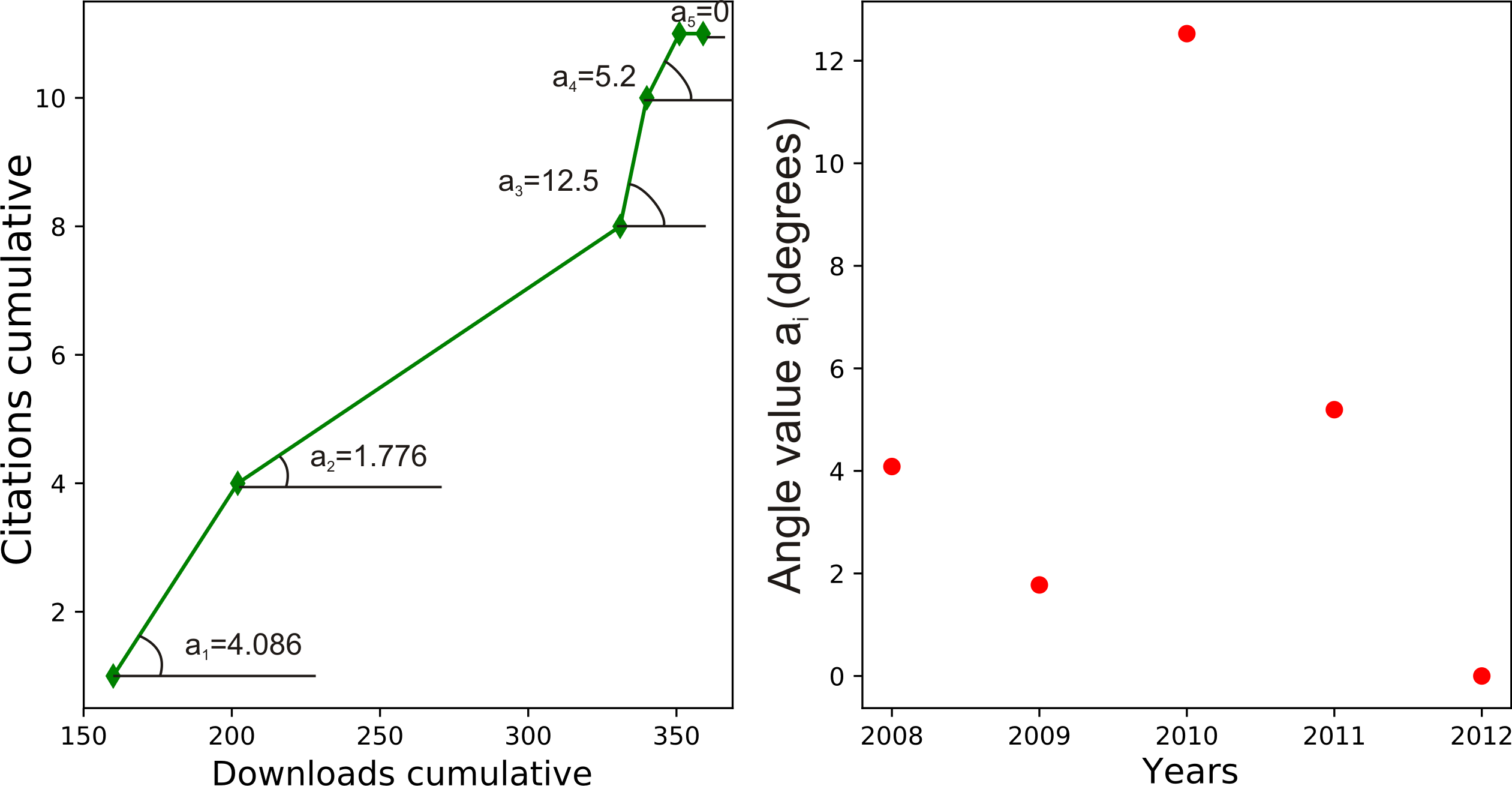}}
\caption{Right-hand panel: Cumulative annual number of citations $c$ vs. cumulative annual number of downloads $d$ for selected paper published in 2007. Left-hand panel: the corresponding values of angles characterizing the annual changes of cumulative curve direction.}
\label{fig_example_zigzag}
\end{figure}

\begin{figure}[!ht]
\vspace{-7mm}
\includegraphics[width=0.5\textwidth]{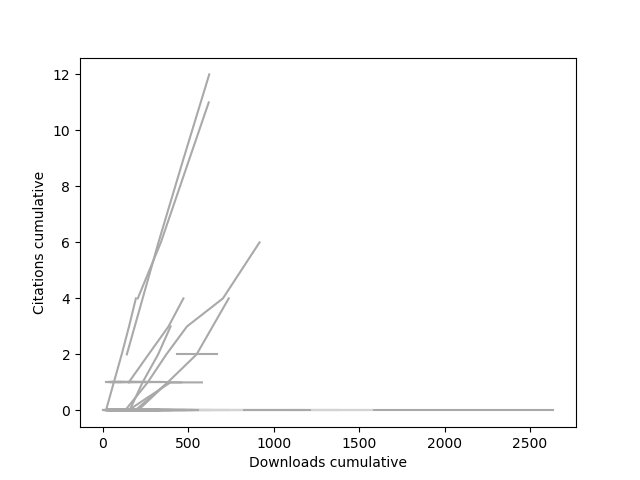}
\\
\includegraphics[width=0.5\textwidth]{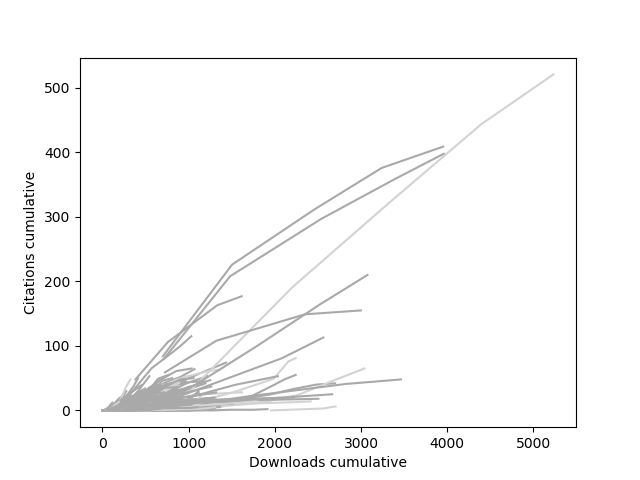}
\\
\includegraphics[width=0.5\textwidth]{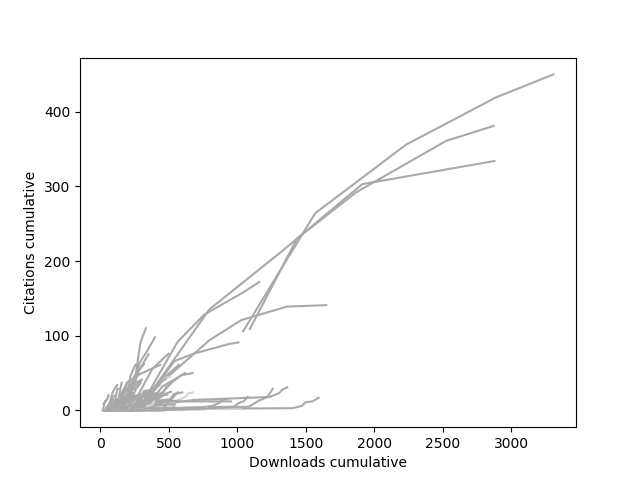}
\caption{Cumulative annual number of citations $c$ vs. cumulative annual number of downloads $d$ for papers with 3 or more years long publication history characterized by $\Delta\leq 0.3$ (top panel), $0.3<\Delta\leq 8$ (middle panel) and $8<\Delta$ (bottom panel).}
\label{fig_example_zigzag}
\end{figure}
The larger is difference between $\Delta=\alpha_{\mathrm{max}}-\alpha_{\mathrm{min}}$ for each paper, the steeper direction change can be found in the corresponding cumulative curve. The empirical investigation of the distribution of  $\Delta$ values for all papers brought us to its possible critical values: $\Delta_{\mathrm{c1}}\approx 0.3$ and $\Delta_{\mathrm{c2}}\approx 8$. Such a criterium to cluster 2\,315 \emph{EPL} papers with 3 or more years long publication history was applied. Three groups of papers are shown in Fig.~\ref{fig_example_zigzag}: 405 papers characterized by slow zig-zag-ness, $\Delta_i\leq\Delta_{\mathrm{c1}}$; 1606  --- medium zig-zag-ness, $\Delta_{\mathrm{c2}}<\Delta_i\leq\Delta_{\mathrm{c2}}$; 304 --- high zig-zag-ness, $\Delta_{\mathrm{c2}}< \Delta_i$.
The clustering of papers into several groups makes further conclusions less reliable statistically. However, while the publication years and kinds of publications are presented more or less similarly in all three clusters of papers, the correlations between cumulative values of downloads $d$ and citations $c$ vary. The smallest value of $\mathrm{PCC}\approx0.19$ is found for the first group. $\mathrm{PCC}\approx0.68$ describes much stronger positive correlation for the second. $\mathrm{PCC}\approx0.84$ for the third category. Therefore, one can state that the strongest correlation between the cumulative counts of downloads and citations was observed for the papers characterized by zig-zag-like cumulative evolution of $c(d)$, i.e. by changing of motives governing the downloading and citing.

\section{Conclusions and outlook} \label{IV}

A case study of downloading and citing of academic publications is presented in this paper. The data from \emph{EPL} journal were used to analyse the dynamics of download process on a scale of entire journal as well as on a scale of individual publications. The first part of the paper contains our recent results on this topic \cite{EPL}. The download rate is naturally different for publications of different age --- the highest interest of potential readers is attracted by newly published papers. Older papers are downloaded less actively. The corresponding cumulative curve of downloads can be considered as one of the characteristics which describes the entire journal. One of the models designed to describe change of users' motives is applied: the function of two exponents allows one to make the long-term predictions about downloading journal publications. The second part of the papers is devoted to the comparison of downloads and citations --- this study is on its initial stage but some very new results are presented here. Not only the correlation between cumulative downloads and citations counts is checked, but a way to cluster publications by their history of downloading and citing is suggested. A further investigation is needed in order to find the reasons for different relationships between these two characteristics.

\section*{Acknowledgment}
We thank {Daniel Barrett (IOP Publishing) and the staff of {\emph{EPL}}} for providing the data and assistance. The work was supported by the grant of the National Academy of Sciences of Ukraine, No.~0118U003620 (O.M.) and by the 7th FP, IRSES project No. 612707 ``Dynamics of and in Complex Systems'' (DIONICOS).

\end{document}